\documentclass{PoS}
\usepackage{gensymb}
\usepackage{graphicx}
\usepackage{url}

\newcommand{\U}[1]{\ensuremath{\mathrm{\ #1}}}
\newcommand{\UU}[2]{\ensuremath{\mathrm{\ #1^{#2}}}}

\title{VERITAS Observations of Fast Radio Bursts}

\ShortTitle{VERITAS Observations of FRBs}

\author{\speaker{Jamie Holder}\\
        Department of Physics and Astronomy and the Bartol Research Institute, University of Delaware, Newark, DE 19713.\\
        E-mail: \email{jholder@physics.udel.edu}}

\author{for the VERITAS Collaboration
        \footnote{for collaboration list see PoS(ICRC2019)1177}}

\author{Ryan S. Lynch\\
        PO Box 2, Green Bank, WV 24944}

\abstract{Fast radio bursts (FRBs) are bright, unresolved, millisecond-duration flashes of radio emission originating from outside of the Milky Way. The origin of these mysterious outbursts is unknown, but their high luminosity, high dispersion measure and short duration requires an extreme, high-energy, astrophysical process. The majority of FRBs have been discovered as single events which would require a chance coincidence for contemporaneous multiwavelength observations. However, two have been observed to repeat: FRB\,121102 and the recently detected FRB\,180814.J0422+73. These repeating FRBs have allowed for targeted observations by a number of different instruments, including VERITAS. We present the VERITAS FRB observing program and the results of these observations.}

\FullConference{36th International Cosmic Ray Conference -ICRC2019-\\
		July 24th - August 1st, 2019\\
		Madison, WI, U.S.A.}

\begin{document}

\section{Introduction}
The phenomenon of fast radio bursts (FRBs) \cite{Lorimer} has been extensively reviewed recently in \cite{Petroff_Review} and \cite{Cordes_Review}. These millisecond-duration events occur at the rate of thousands per day and are now known to be extragalactic in origin, implying radio flux densities $\sim10$ billion times larger than Galactic pulsars. The nature of the sources is unclear; the bursts show a wide range of different properties in the radio observations, and many different emission models have been proposed (as discussed in the aforementioned reviews). Two FRB sources have now been shown to repeat: FRB\,121102 \cite{FRB121102} and FRB\,180814.J0422+73 \cite{FRBJ0422}. For FRB\,121102, radio polarization measurements indicate an extreme magneto-ionic environment at the source \cite{FRB121102_pol}. In this case, a rapidly rotating young magnetar, or a neutron star in the immediate environment of a massive black hole, provide a plausible explanation, but alternative models still exist.  

The discovery of prompt counterparts at any wavelength would strongly discriminate between the various possible origin scenarios and allow an estimate of the bolometric luminosity of FRBs. Space-based searches for high energy FRB counterparts have been performed using both instruments on board the Fermi Gamma-ray Space Telescope \cite{GBM_FRBsearch, LAT_FRBJ0422}. Imaging atmospheric Cherenkov telescopes (IACTs) are uniquely suited to conduct such searches from the ground. At GeV-TeV energies, IACTs provide an instantaneous effective area of $\sim10^5\UU{m}{2}$; five orders of magnitude greater than space-based detectors. For millisecond-timescale phenomena, they are essentially background free. In addition, IACTs are the largest \textit{optical} telescopes in the world. Their spatial resolution is very poor, by the standards of optical astronomy, with typical optical point spread functions of a few arc minutes; however, with mirror areas $>100\UU{m}{2}$, and with multi-telescope arrays equipped with fast photosensors, they are ideal instruments for optical photometry of very rapid transient sources (e.g. \cite{OSETI, Asteroid}).

In searching for optical or gamma-ray counterparts to FRBs with IACTs, the difficulty is knowing when and where to look. The field of view of IACTs is a few degrees in diameter and slewing the telescopes to the correct location takes seconds or minutes. Until recently, observations of FRB locations with IACTs were therefore limited to the search for afterglow emission \cite{HESS_FRB}. The discovery of repeating FRBs allows contemporaneous monitoring of FRB locations with IACTs and radio telescopes, a technique which has already been exploited to place limits on gamma-ray and optical emission from five bursts of FRB\,121102 with the MAGIC telescopes \cite{MAGIC_FRB}. In this contribution, we report on the status of searches for optical and high energy emission from FRBs with the VERITAS IACT array.

\section{VERITAS}
VERITAS is an array of four, $12\U{m}$ diameter atmospheric Cherenkov telescopes located at the Fred Lawrence Whipple Observatory in Arizona \cite{Holder06,Park15}. The telescopes are located on the corners of an approximate square, with side lengths of $\sim100\U{m}$. Each steerable telescope is equipped with a 499-pixel photomultiplier tube camera, covering a circular field-of-view with a diameter of $3.5\degree$. For observations in the gamma-ray band, images of the Cherenkov light from cosmic ray and gamma-ray initiated air showers are recorded and processed using standard tools \cite{Maier17,Daniel_VEGAS}. Digitization of the AC-coupled photomultiplier tube signals occurs at 500 megasamples-per-second, for Cherenkov images which exceed the array trigger thresholds. The readout is restricted to a 32\U{ns} window around the time of the trigger. 

The requirements for millisecond-timescale optical photometry are significantly different to those of gamma-ray observations, and so a specialized data acquisition chain is used which is fully parallel to normal Cherenkov operations. Details are given elsewhere in these proceedings \cite{Daniel_ECM} but, briefly, the DC-coupled photomultiplier tube voltages are recorded using a commercial 14-bit data logger (DATAQ instruments DI-710-ELS) on up to 16 channels for each telescope. The system samples at a rate of $4,800/n\U{Hz}$, where $n$ is the number of channels read out, and has a sensitivity roughly equivalent to a limiting magnitude of 11 for an astronomical B-filter.

\section{Observations}

Results of a search for steady gamma-ray emission from the location of FRB\,121102 with 10.8 hours of VERITAS observations taken over 2016-2017 have been reported in \cite{Bird_ICRC}. While some of these exposures were coincident with Arecibo observations, no radio bursts were reported during the overlapping observations. Additional VERITAS observations of FRB\,121102 on 25 November 2017 (MJD 58082) were made in coincidence with the Robert C. Byrd Green Bank Telescope (GBT). Data were recorded using the Green Bank Ultimate Pulsar Processing Instrument (GUPPI) with a bandwidth of 800 MHz centered on a radio frequency of 2000 MHz.  Full-polarization spectra were recorded every $10.24\U{\mu s}$ with a frequency resolution of $1.5625\U{MHz}$ and coherently dedispersed at a dispersion measure of $560\U{pc} \UU{cm}{-3}$.  The data were searched for bright pulses from dispersion measures of 527 to $587\U{pc}\UU{cm}{-3}$ using the \texttt{single\_pulse\_search.py} tool from the \texttt{PRESTO}\footnote{\url{https://www.cv.nrao.edu/~sransom/presto/}} software suite.  Spectrograms were created and inspected for all candidates with signal-to-noise ratio > 7 in order to differentiate between astrophysical sources and terrestrial sources of radio frequency interference.  During 115 minutes of VERITAS observations (from MJD 58082.413 to 58082.495), 17 radio bursts were detected by GBT, 2 of which occurred between VERITAS observing runs when the telescopes were not acquiring gamma-ray data. The remaining 15 bursts have contemporaneous optical and gamma-ray exposures with VERITAS, and are presented here. Exact burst times will be reported in a future publication.

VERITAS observations of the more recently identified repeating FRB,  FRB\,180814.J0422+73, were conducted between 29 December 2018 and 12 February 2019. The exposures total 8.2 hours and were timed to provide coincident observing with the Canadian Hydrogen Intensity Mapping Experiment (CHIME) \cite{CHIME_Instrument}. No radio bursts have been reported during the VERITAS observations.

\section{Results}
\subsection{Gamma-ray search}
\begin{figure}[ht!]
\centering
\includegraphics[scale=0.7]{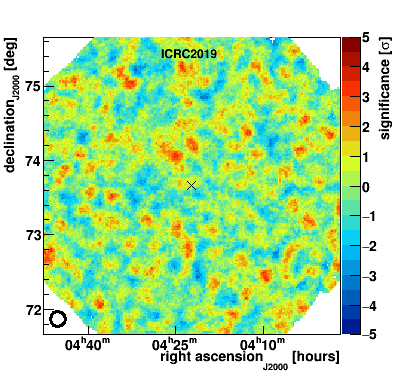}
\caption{Significance skymap of the region around FRB\,180814.J0422+73. The "$\times$" indicates the location of the repeater, while the black circle in the lower left shows the approximate size of the gamma-ray point spread function. No evidence for steady gamma-ray emission is seen.}
\label{J0422_skymap}
\end{figure}

No evidence for steady gamma-ray emission has been seen from either of the repeating burst locations. Two analyses were performed, with \textit{soft} and \textit{moderate} gamma-ray selection cuts using boosted decision trees trained on gamma-ray simulations and real cosmic ray background \cite{BDTs}. These cuts correspond to the lowest energy threshold and to more strict background rejection, respectively. The 95\% confidence integral flux upper limits for FRB\,180814.J0422+73 are $9.24\times10^{-13}\U{photons}\UU{cm}{-2}\UU{s}{-1}$ above $300\U{GeV}$ for \textit{soft} cuts and $6.26\times10^{-13}\U{photons}\UU{cm}{-2}\UU{s}{-1}$ above $500\U{GeV}$ for \textit{moderate} cuts, for an assumed $E^{-2}$ differential power law spectrum. The relatively high energy thresholds are a result of the moderately low elevation angle at which these observations were made ($\sim46\degree$). Figure~\ref{J0422_skymap} shows the significance skymap for observations of FRB\,180814.J0422+73 using \textit{moderate} cuts. 

Upper limits for the archival observations of FRB\,121102 have been reported in \cite{Bird_ICRC}. The 95\% confidence limits for steady emission during the observations of 25 November 2017 alone, during which 15 FRBs occurred, are $5.25\times10^{-13}\U{photons}\UU{cm}{-2}\UU{s}{-1}$ above $200\U{GeV}$ for \textit{soft} cuts and $1.86\times10^{-12}\U{photons}\UU{cm}{-2}\UU{s}{-1}$ above $300\U{GeV}$ for \textit{moderate} cuts.

The methodology for a comprehensive burst search has been discussed in \cite{Bird_ICRC} and \cite{MAGIC_FRB}. It typically includes searching multiple time windows, with multiple levels of event selection, as well as searching for delayed or precursor emission. This complete analysis will be reported in a future publication. Here we present the simplest analysis only --- a search for strictly contemporaneous gamma-ray and radio emission from FRB\,121102. 
After the application of standard \textit{soft} gamma-ray selection cuts, the mean number of background events within a $10-$ms window centered on the reported radio burst time is $9\times10^{-5}\U{events}$, where the background is estimated using the standard \textit{ring-background} method \cite{Berge}. At the location of FRB\,121102, no gamma-ray like events passed the cuts for any of the 15 bursts with contemporaneous observations. This is consistent with expectation in the absence of a signal (the probability of observing zero events is 0.99991 for each burst). Following \cite{MAGIC_FRB}, since the background rate is negligible, the 95\% confidence upper limit to the number of events per burst is 3.56 events, corresponding to an integral flux upper limit of 
$5.6\times10^{-7}\U{photons}\UU{cm}{-2}\UU{s}{-1}$ above $200\U{GeV}$. The combined upper limit over all 15 bursts is $3.7\times10^{-8}\U{photons}\UU{cm}{-2}\UU{s}{-1}$.

\subsection{Optical search}

During the observations of 25 November 2017, the VERITAS optical photometry hardware was only partially commissioned. The system was installed on only one of the telescopes (Telescope 1), and 12 photomultiplier tubes channels were connected, including the central pixel which was observing the location of FRB\,121102. The sampling rate was $300\U{Hz}$. The data were not accurately timestamped --- although accurate timing may be recoverable by comparing the optical photometry to bright transient optical events (meteors) recorded in the gamma-ray data acquisition path, which includes GPS timing.

\begin{figure}[ht!]
\centering
\includegraphics[width=\textwidth]{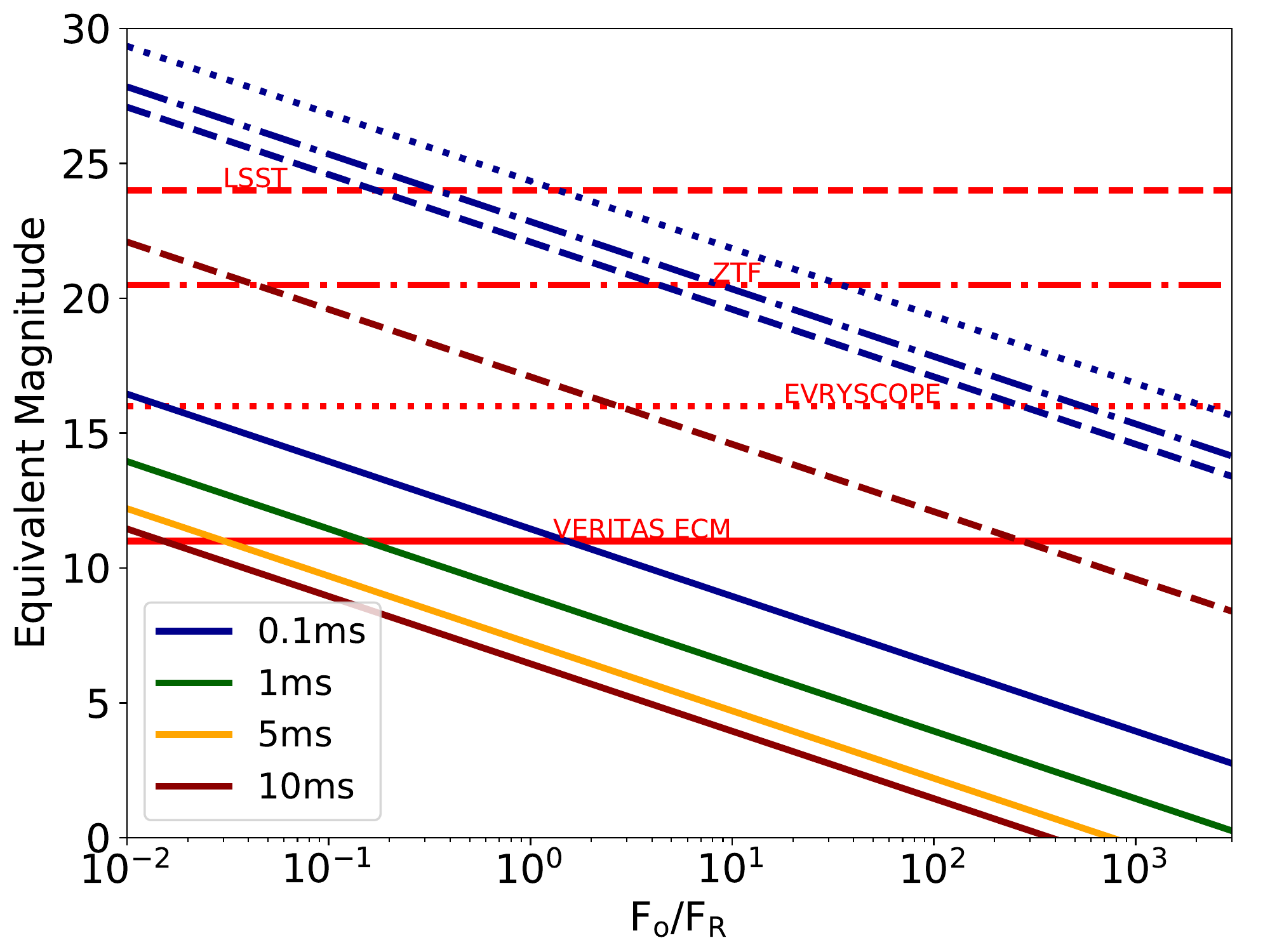}
\caption{The equivalent magnitude for an optical counterpart to an FRB as a function of the radio luminosity, following the scheme of \cite{Lyutikov2016} for an FRB of 0.4\,Jy and optical burst durations of 0.1 (blue), 1 (green), 5 (yellow) and 10\,ms (burgundy) respectively. Solid lines are for an observation exposure of 1\,ms (VERITAS ECM), 15\,s (LSST), 30\,s (Zwicky Transient Factory) and 120\,s (EVRYSCOPE). The red lines are the limiting magnitudes for the respective instruments.}
\label{fig:optical}
\end{figure}

A preliminary search for optical signals of <0.1\,s at signal/noise of $>10$ is consistent with only background events (likely meteors). Plotted in figure~\ref{fig:optical} are the expected magnitudes for optical counterpart bursts of duration from 0.1 to 10\,ms following the scheme of \cite{Lyutikov2016}. 
The VERITAS limits range from $\sim 1$\% of the radio energy flux for an assumed 0.4\,Jy radio burst with a 10\,ms optical burst duration, up to a matching radio flux for an optical counterpart of 0.1\,ms duration. 
Also plotted is a comparison of the VERITAS optical photometry sensitivity compared to a number of optical transient facilities: the Large Synoptic Sky Telescope (LSST), the Zwicky Transient Factory (ZFT), and the EVRYSCOPE. 
The VERITAS limits are within a factor of 10 of those achievable by the LSST and superior to many other transient facilities due to their long ($>15$\,s) exposures compared to the short burst duration, demonstrating that VERITAS is an excellent facility for judging the optical counterpart fluence for FRBs.

\section{Discussion}
The search for fast radio burst counterparts using imaging atmospheric Cherenkov telescopes is still in its earliest stages. Coordinated observing campaigns on the repeating burst locations will continue, but the prospects to observe non-repeating bursts are also extremely promising and may probe an entirely different class of object \cite{FRB190523}. While fewer than 100 bursts detections have been published so far, the extrapolated event rate across the sky is thousands per day. IACT archives, therefore, already contain many observations of FRBs, unobserved by any radio telescope --- approximately one for every few days of observations, depending on the IACT field of view and the true rate of FRBs. A search of these extensive archives for gamma-ray event clusters on millisecond timescales is therefore worthwhile. Previous "blind" burst searches have been performed, notably those motivated by predicted emission from evaporating primordial black holes, but these have typically used a lower limit of seconds, as opposed to milliseconds \cite{HESS_PBH, VERITAS_PBH}. 

The commissioning of CHIME provides an even more promising avenue to explore, and one for which VERITAS is uniquely suited. CHIME is now operating and is continuously monitoring a $\ge 200\mathrm{deg}^2$ overhead strip of the sky. VERITAS and CHIME both view the northern sky and are separated in longitude by $8.7\degree$. This allows contemporaneous radio and IACT searches for non-repeating FRBs with a high probability of successfully performing a coincident observation. In order to overlap VERITAS observations with the CHIME field of view, it is simply necessary to track a position on the celestial sphere 35 minutes after culmination, a requirement which is regularly met during standard gamma-ray observations and can easily be factored into the VERITAS observing program. Coincident CHIME/VERITAS observations have been performed throughout 2019 and will be examined for overlapping burst observations once the CHIME results are made public.

Optical photometry of FRBs with IACTs also has a promising future. All four VERITAS telescopes are now equipped with the necessary electronics and typically operate with 2 photomultiplier tube channels sampled at $2400\U{Hz}$. The four-telescope VERITAS array is particularly well-suited for such searches, since the telescope separation of $\sim100\U{m}$ leads to a noticeable parallax of half of the photomultiplier tube pixel diameter ($0.07\degree$) for meteors at a typical altitude of $\sim80\U{km}$. Astrophysical optical transients will not be subject to this effect, and so comparing the optical intensity and pulse profile between multiple telescopes should allow the otherwise dominant local background to be suppressed.

\section{Acknowledgements}
This research is supported by grants from the U.S. Department of Energy Office of Science, the U.S. National Science Foundation and the Smithsonian Institution, and by NSERC in Canada. This research used resources provided by the Open Science Grid, which is supported by the National Science Foundation and the U.S. Department of Energy's Office of Science, and resources of the National Energy Research Scientific Computing Center (NERSC), a U.S. Department of Energy Office of Science User Facility operated under Contract No. DE-AC02-05CH11231. We acknowledge the excellent work of the technical support staff at the Fred Lawrence Whipple Observatory and at the collaborating institutions in the construction and operation of the instrument.

\end{document}